\documentclass{article}

\usepackage[pdftex]{graphicx}
\usepackage[preprint]{neurips_2019}

\usepackage[utf8]{inputenc} 
\usepackage[T1]{fontenc}    
\usepackage{hyperref}       
\usepackage{url}            
\usepackage{booktabs}       
\usepackage{amsfonts}       
\usepackage{nicefrac}       
\usepackage{microtype}      

\title{Studying Shallow and Deep Convolutional Neural Networks as Learned Numerical
  Schemes on the 1D
  Heat Equation and Burgers' Equation}

\author{%
  Alejandro Francisco Queiruga\\
  Earth and Environmental Sciences Area\\
  Lawrence Berkeley National Lab\\
  Berkeley, CA 94720\\
  \texttt{afqueiruga@lbl.gov}
}

\begin{document}

\maketitle

\begin{abstract}
  This paper examines the coincidence of neural networks with
  numerical methods for solving spatiotemporal physical
  problems. Neural networks are used to learn predictive numerical
  models from trajectory datasets from two well understood 1D
  problems: the heat equation and the inviscid
  Burgers' equation. Coincidence with established numerical methods is shown by demonstrating that a
  single layer convolutional neural network (CNN) converges to a traditional finite difference stencil for the
  heat equation. However, a discriminator-based adversarial training
  method, such as those used in generative adversarial networks (GANs), does not find the
  expected weights. A compact deep CNN is applied to nonlinear Burgers'
  equation, where the models' architecture is reminiscent of existing winding
  finite volume methods.
  By searching over architectures and using multiple
  recurrent steps in the training loss, a model is
  found that can integrate in time, recurring on its outputs, with
  similar accuracy and stability to Godunov's method.
\end{abstract}

\section{Introduction}

The physical systems at the limits of forecasting capabilities are challenging
due to a combination of unknown underlying physics and
traditional approaches being computationally intractable. Data-driven analysis of dynamics 
through neural networks and deep learning is a promising approach and
a hot topic, but the properties of the methods are not yet well understood. The problem of discovering dynamics can be stated as follows:\footnote{This approach seeks a
  function that maps an image to an image. Another approach is to look
  for conditional scalar functions with coordinates as inputs,
  $u^{k+1}(x,y)=f(x,y|u^{k})$; this was not considered here. }
\begin{equation}
\textrm{Given data }\, u_i^k=u(x_i,t_k), \,\textrm{find }f\textrm{ such that }\, u^{k+1}=f(u^k)
\end{equation}
where $u$ is the physical observable, $k$ is a time index and $i$ is a space index. The use of artificial neural networks
(ANNs) as an $f$ is explored in this paper to
discover predictive functions given data from well known partial differential
equations (PDEs), for which
decent $f$s are already known from the history of numerical analysis.

This paper takes the viewpoint that the use of ANNs directly
searches for a numerical operator, as opposed to fitting to features
derived from PDEs.
ANNs are rapidly being applied to physical systems; for example, long
short-term memory networks \citep{vlachas_data-driven_2018}
and GANs are
being applied to physical problems \citep{xie_tempogan:_2018,wu_physics-informed_2018,werhahn_multi-pass_2019} . 
Problems in the physical sciences require fine-grained properties such as regression accuracy and numerical stability.
It has been suggested that the structure of CNNs and not the
exposure to datasets is the dominating factor to their performance
\citep{ulyanov_deep_2018,zador_critique_2019}. Thus, carefully
checking existing methods is warranted, but, on the other hand, devising
architectures specifically for physical applications will potentially be fruitful.

Two well-understood 1D time-dependent problems are treated:
the heat equation, $u_{t} = k u_{xx}$, 
and Burgers' equation,  $u_{t} + u_x u = 0$. The heat equation is linear and has a 
known finite difference stencil. Burgers' equation, however, is very
nonlinear and even yields discontinuous solutions. Many complex numerical schemes for Burgers' equation
exist with various success.
Even this 1D equation is still an open problem where data-driven
approaches can be applied; the recent work of
\citet{bar-sinai_data-driven_2018} successfully learned high-order
reconstructions of the fluxes from high-resolution simulated data of the viscous
Burgers' equation.

\begin{figure}
  \centering
  \includegraphics[width=5in]{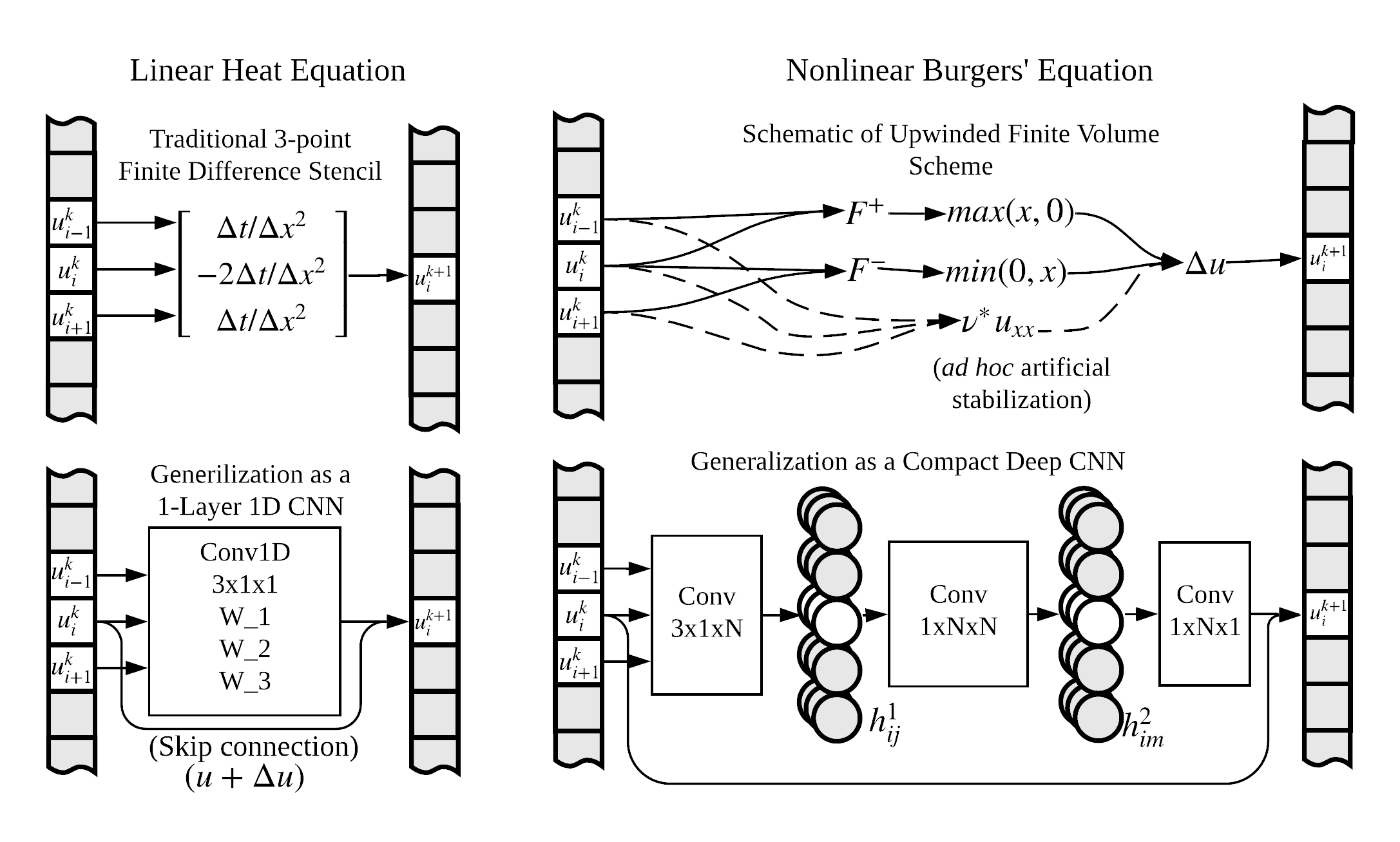}  
  \caption{\label{fig:cnn_fdm}Coincidence of CNNs with existing finite difference
    schemes. On the left, a one-layer CNN is the same as the traditional 3-point finite difference
    stencil. It is demonstrated that the 3 weights converge to the
    expected parameters. On the left, a finite volume scheme for
    Burgers' equation requires a complex nonlinear graph with
    decision trees for winding and sometimes nonphysical stabilization terms. A generalized CNN can attempt to
    discover a similar, or even better, method. Including the skip connection
    allows the model to be generalized to other timestep sizes, and
    might improve recurrent training.}
\end{figure}

As illustrated in Figure \ref{fig:cnn_fdm}, some numerical schemes
can viewed as a fringe case of certain CNN architecture. The finite
difference method uses Taylor expansions to derive update rules for the next time step:
\begin{equation}
  u^{k+1}_{i} = u^k_{i}+\mathtt{stencil}\left( u^k_{i-1}, u^k_{i},u^{k}_{i+1}; \Delta x, \Delta t\right)
\end{equation}
The classical stencil for the heat equation is
$u^{k+1}\approx u^k+(k\Delta t)/(\Delta x^2) \left(u_{i-1} - 2 u_{i} +
  u_{i+1}\right)$.
This architecture corresponds to a fringe case ANN: a
3-weight 1D convolutional neural network (CNN) with no bias and no
activation function. Verifying that
these coefficients can be derived
by the learning strategy and optimization algorithm is proposed as a good first step.

Forecasting far into the future is of interest. As a standard approach
in numerical methods, it is desired for a model to be able to recur on
its own outputs without lossing accuracy or stability:
$u^{k+n}=f(f(...f(u^k)$. The connection between numerical integration
and recurrence is in active study, with analogies to ANNs made by
\citet{chen_neural_2018} and \citet{chang_multi-level_2017}.

The 1D problems were specifically chosen to yield quickly reproducible
experiments. For each equation, a dataset
including different trajectories from different initial conditions is made
using analytical solutions with Sympy. These are evaluated on a grid of 41 points in $x$ and 100
snapshots in $t$, for a total size of $\approx$1.6MB. Each experiment runs
in a few minutes using a GPU. The
entire study is implemented in PyTorch. The source code, datasets,
and figures for this study can be found at https://github.com/{\em omitted for blind review.}

\section{The Heat Equation}

The first step to test the methodology is to derive the $[1,-2,1]$
stencil from a dataset of heat equation trajectories with a three-parameter CNN.
The dataset contains ten trajectories for various trigonometric and polynomial
initial conditions with $u=0$ boundary conditions computed with the
Fourrier series analytical solution. Two trajectories each are used for
testing and validation.
To ensure stability for an explicit scheme, the domain was $x\in[0,1]$,
$t\in[0,1/4]$ and the diffusion coefficient was
$k=1/10$, informed by the Courant–Friedrichs–Lewy (CFL) condition $\delta t\leq k \delta
x^2$\citep{leveque_finite_2007}.\footnote{As with implicit numerical methods, a
  different ANN architecture may be able to surpass this condition.}

A combination of standard training using the mean squared error (MSE) and
adversarial training with a discriminator is considered. A conditional
discriminator $D(y|x)$ is optimized which learns, given $x$, to determine
if $y$ is the datum or the model prediction. For these problems, no
stochastic effects are included, and the model and evaluation of the
discriminator are deterministic. Thus, the
discriminator essentially {\em learns a loss function}, replacing the
mean-squared-error loss with potentially something better:
\begin{equation}
L\left(u^k,u^{k+1}\right) = \lambda_1 \left\| u^{k+1}-f(u^k)
\right\|_2^2 + \lambda_2\left(1 - D\left(f(u^k)|u^k\right) + D\left(u^{k+1}|u^k\right)\right)/2
\end{equation}
The cost function is the mean of the loss function over the batch.

The weights $\lambda_1$ and $\lambda_2$ are set to $(1,0)$, $(0,1)$, and $(1,1)$. When $\lambda_2\neq 1$, the
discriminator $D$ is trained to maximize the cost function alternating
steps with the model $f$.

\begin{table}
  \caption{\label{tab:conv}The convolutional weights learned for the
    heat equation for five randomly initialized runs with three
    training strategies. The weights are
    divided by $k \Delta t / \Delta x^2$ to normalize to
    $1,-2,1$. Training through a discriminator does not get the
    correct magnitude, but, interestingly, learns the shape.}
  \centering
  \begin{tabular}{rrr|rrr|rrr}
    \hline
    \multicolumn{3}{c}{MSE} & \multicolumn{3}{c}{Discriminator Adversary}  &\multicolumn{3}{c}{Both} \\
\hline
 0.997 & -1.995 & 0.997 & 1.398 & -1.149 & 1.401 & 1.073 & -2.149 & 1.077 \\
 0.997 & -1.995 & 0.998 & 1.058 & -3.196 & 0.956 & 0.994 & -2.000 & 0.995 \\
 0.998 & -1.996 & 0.998 & 1.512 & -0.824 & 1.624 & 0.595 & -1.502 & 0.732 \\
 0.998 & -1.994 & 0.999 & 1.084 & -1.154 & 1.116 & 1.000 & -2.001 & 1.000 \\
 0.998 & -1.995 & 0.998 & 1.392 & -0.558 & 1.394 & 1.000 & -1.999 & 1.000 \\
\hline
\end{tabular}
\end{table}

The experiment was repeated 5
times for each loss function and the weights are reported in
Table \ref{tab:conv}.
The MSE loss achieves $10^{-7}$ error, which is likely the best obtainable in single precision.
Purely adversarial training with a discriminator continued for ten
times as many epochs and does not
learn the same stencil. It appears that the discriminator only learns
the shape, but not the magnitude. Combining the
discriminator loss and L2 loss did not succeed every time and
converged slower (in number of steps). 
The architecture of the discriminator was a pooling CNN with three
hidden layers and LeakyReLU activation functions, with a total of 51
parameters. Its architecture should be more thoroughly
studied to make a firm conclusion. Increasing from single
precision to double precision did not change the results.

\section{Burgers' Equation}

The dataset contained 20 trajectories with a series of linear profiles, shock and rarifaction
profiles of the Riemann problem, and one parabolic
profile. Anti-reflections were included to encode the symmetry.
The CFL equation for this equation is $\Delta t < C \Delta
x / \max lu|$, so the domain was set to $x\in[-1,1]$, $t\in[0,1]$, and
the velocities were kept below 2.

The compact deep CNN architecture has the following three hyperparameters: the number of
features in the hidden layers $n$,the total depth of the network $d$,
and the activation function, $\sigma$. The layering of the
architecture is: 
Conv(1,$n$,3), $\sigma$,Conv($n$,$n$,1)... $d-1$ times... $\sigma$,
Conv($n$,1,1).
The following activation functions were tried: ReLU, LeakyReLU, Tanh,
CELU, Sigmoid. The depth was varied from 2-4, and number of channels
from the set 3,5,10, and 15.
The width of the first convolution, 3, was not varied in this study,
but is under active research.

\begin{figure}
  \centering
  \includegraphics[width=2.75in]{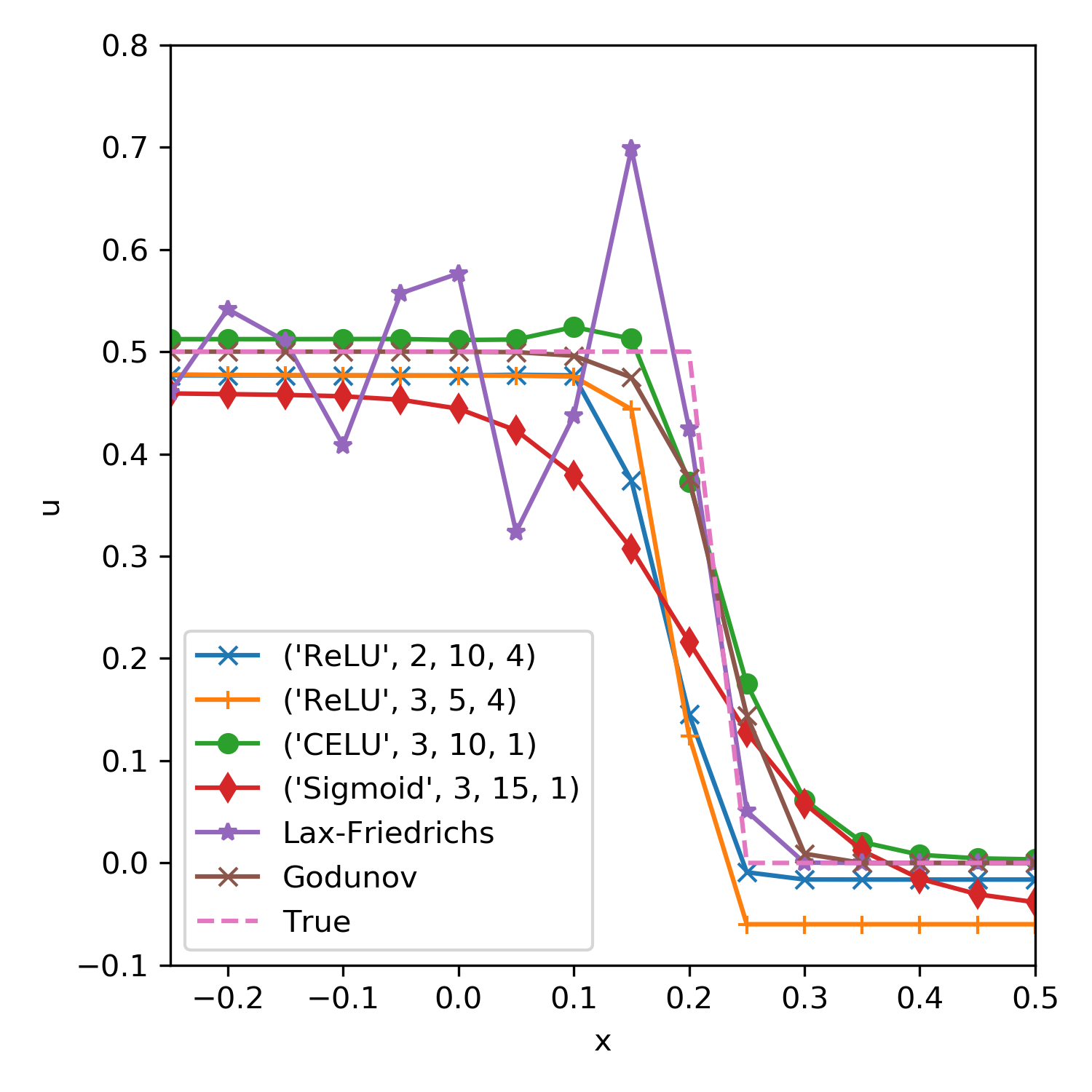}%
  \includegraphics[width=2.75in]{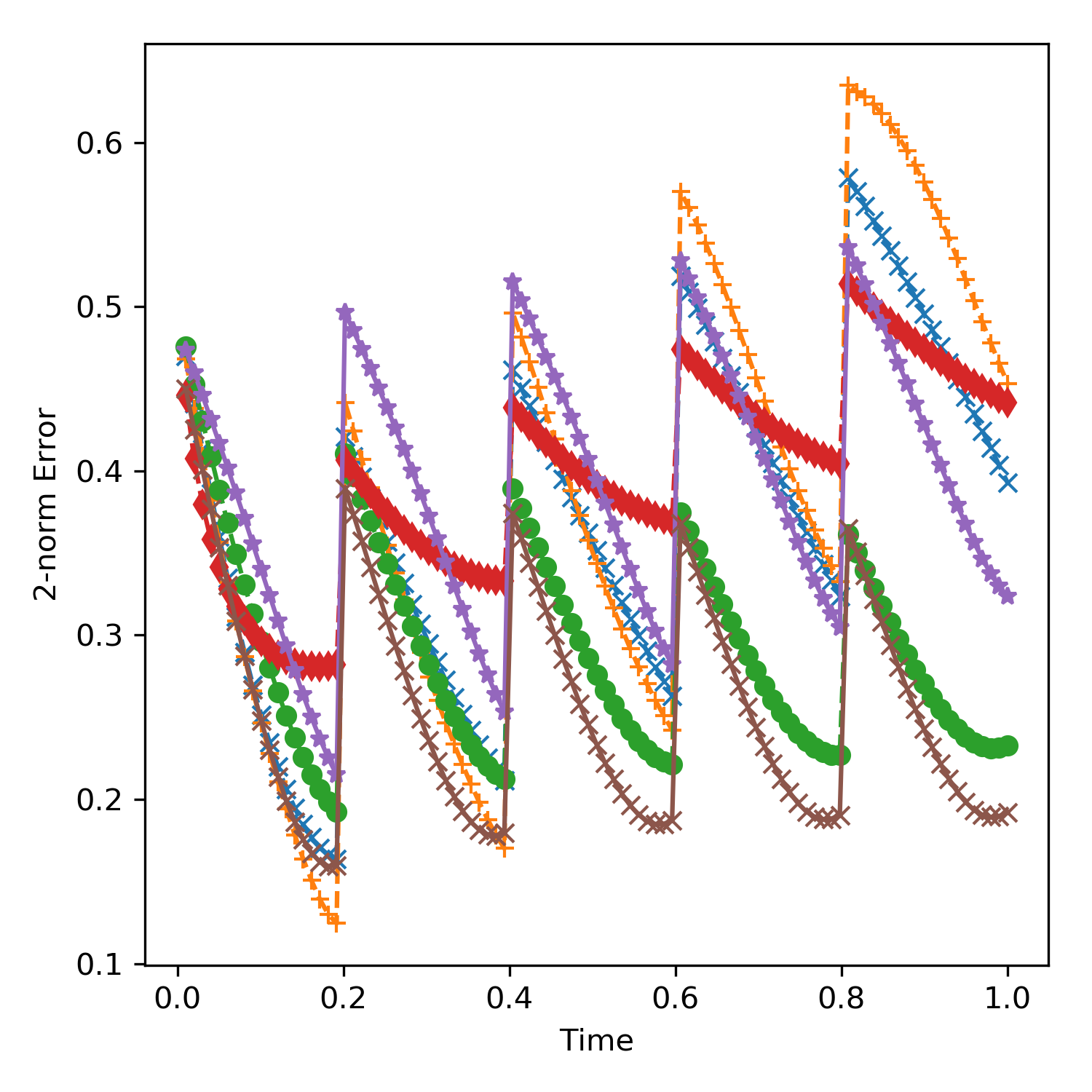}
  \caption{\label{fig:shock_comparison}Performance of the best learned models against well
    developed methods on a shock that did not appear in the
    training set. The legends are labeled by (activation, depth,
    channels, and terms in Eq. 4). Left, a snapshot of the methods with the true
    solution after 100 steps to $t=1$. (To stress: the CNNs recurred on
    themselves 100 times.) Right, error of the
    methods compared to the analytical solution over time. }
\end{figure}

To improve training with recurrent prediction as a goal,
multiple steps were included in the training:
\begin{eqnarray}
L\left(u^k,\left\{u^{k+1},u^{k+2},u^{k+3}...\right\}\right) =  \left\|
  u^{k+1}-f(u^k) \right\| + \lambda_1  \left\| u^{k+2}-f \circ f(u^k)
  \right\| ...\nonumber
  \\
   + \lambda_2  \left\| u^{k+3}-f \circ f \circ f(u^k) \right\| + ...
\end{eqnarray}
where $\lambda_i$ are weighting coefficients that were set to one in
this case. Hyperparameter variation using 1,2,3, and 4 steps
qualitatively demonstrated an
improvement in overall stability of the models. The search space included 180 different networks and loss
function combinations. Learning a discriminator did not have a positive effect on the
results for this problem.

The learned model is compared to implementations of the classical
numerical schemes of Lax-Friedrichs and Godunov. (See
\citet{leveque_finite_2007} and  \citet{godunov_difference_1959}.) This
profile did not appear in either training or testing. The final
profile and error across is shown in Figure
\ref{fig:shock_comparison}. 
The best learned model is less accurate than Godunov's, but performs
similarly. The Lax-Friedrichs method exhibits instability, a well
known-phenomenon. This behavior was seen in other model
architectures on other problems not shown. Further demonstrations with more model
architectures can be found online.

\section{Conclusion}
\label{sec:conclusion}

We show that a fringe case of CNN architecture corresponds to a
standard finite difference stencil, and converges to the expected
coefficients using popular optimizers for ANNs on the L2 loss but not with a
learned loss function through adversarial training.
These results suggest caution when using a purely GAN-type training for physics problems where accuracy is important. The ability to detect the shape of the operator is promising; the {\em author(s)} hypothesize that the discriminator may help with issues such as stability in more complex systems.
Deep CNNs were successfully learned for solving Burgers' equation accurately and stably.
By searching for compact models on small solutions, the model can be applied to domains
with different geometries.

Applying intuition from well understood physics-and-math-up approaches
will improve future approaches, providing insights that can hopefully
be applied to problems without known physical descriptions but
similarities to canonical problems. Studying the stability properties of recurring these networks applied
to physics problems can extend to stabilizing recurrent networks for
other applications.
By finding this area of overlap between solving PDEs and deep
learning, we can seek to bridge the gap and transfer knowledge between
the two fields.

\subsubsection*{Acknowledgments}

{\em Placeholder for blind review.}

\bibliographystyle{plainnat}
\bibliography{zotero}

\end{document}